\documentclass[showpacs,aps,preprint,epsfig,nofootinbib]{revtex4-1}
\usepackage{epsfig}
\usepackage{graphicx}
\usepackage{amsfonts}
\usepackage{mathrsfs}
\usepackage{charter}

\date{\today}%

\begin{document}%


\title{The cuprate superconductors: a phenomenological overlook} 

\author{ CHENG Jian-Feng } 
\email{chengjf@cqupt.edu.cn}
  
\affiliation{
 College of Mathematics $\&$ Physics, Chongqing University of
Posts and Telecommunications, Chongqing 400065, China }

\date{\today}
\begin{abstract}
 We, basing on the quantum critical point (QCP)($p = 0.19 $), 
propose a phenomenological description on high-$T_c$ superconductivity
in the cuprate superconductors, and suggest it divides
the whole doping region into two parts: the underdoped region
($p < 0.19 $) and the overdoped region ($ p> 0.19 $). The electrons 
 in the former are localized and form the localized Fermi
liquid, a kind of the non-Fermi liquid, where the carriers
are the holes; the electrons in the latter are itinerant
and form the Fermi liquid, where the carriers are
the electrons.

We further argue that localization is a prerequisite to the pseudogap;
the superconductivity gap and coherence forms at the same time, which both
share the same energy scale; coherence induces phonon around the node.
\end{abstract}

\maketitle

In 1986 J.G. Bednorz and K.A. M$ \ddot{\rm u}$ller discovered high–critical
temperature ($T_c$) superconductivity\cite{BM_exp}, 
which opened a new chapter in strongly correlated system. 
They found an onset of the superconductivity in $ \rm La_{2-x}Ba _x CuO_4 $
near $ 30 ^\circ \rm K $, modest but still 
far above the previous values observed in the conventional metals.
In the following months, it continued to ascend: to $ 45^ \circ \rm K $,
to $ 52 ^ \circ \rm K $, and so far as to reach 
$ 93 ^\circ \rm K $ in $ \rm Y Ba _2 Cu _3 O_{ 7-y } $\cite{CWChu_exp}. 
But astonishment just began to unveil. In the following 
years, experimentlists found a series of fancy features, 
such as pseudogap, stripe and so on. 

The great progress in experiment not only provides a wide
stage for theorists, but also places a heavy burden 
on their shoulders. 
With the push from experiment, the theorists have done 
a great many of works and reached a lot of consensuses, but are still
facing a great challenge. On the front of the experimental
phenomena a bewildering variety of theoretical 
models fall short of our expectations.
A unified interpretation on the phase diagram, for example, 
is still out of control, and even the phase diagram itself
is still in dispute.

What is the cornerstone to support so many strange properties in the
cuprate superconductors? It is difficult, due to the complexity,
to cover all the experimental phenomena at one blow.
We have to curtail the branches and concentrate
ourselves in some extreme doping cases, 
for doping dominates the properties of the cuprate superconductors.

\section{Doping}

Fundamentally different from a conventional (band) insulator, 
the parent compounds of the cuprate superconductors
is a Mott insulator\cite{Anderson1987}. Even it is half-filled,
one electron per unit cell and the other on-site orbit idle, 
it cannot conduct. To a conventional insulator, however, such idle orbit can be 
occupied and used to conduct electricity. It is the strong on-site
electron-electron repulsion that blocks such motion and makes 
the electrons {\it localized}. \normalsize Does such localization
still exist in the doping compounds?

We consider two ultra doping cases: the ultra-low doping case ($ p\sim 2\%$)
and the ultra-high doping case.

With the introduction of very few holes, most of the electrons are still 
localized and only some electrons around holes can 
move by hopping to form conductivity: electrons form a {\it localized Fermi
liquid\normalsize}, a kind of the non-Fermi liquid. The carriers thus are holes
and their concentration is proportional to 
the hole concentration $p$.

In the ultra-high doping case, holes are enough to make electrons
“live” apart, moving with very low mutual interference: electrons
are no longer localized and begin to itinerate, their density of possibility
filling respectively the whole copper oxide plane. That is to say,
the behaviour of electrons is like the Fermi liquid.
The carriers here are not holes but electrons and 
their concentration is proportional to the concentration of electrons
$(1-p)$, not the concentration of holes $p$. 

The above discussion implies the properties of the electrons in two ultra cases 
are fundamentally different, for one is the
non-Fermi liquid and the other 
is the Fermi liquid. What is responsible for the difference? Doping! 
Although we just consider two ultra cases, it is reasonable to speculate
that the doping will lead the cuprate superconductors to evolve from the 
non-Fermi liquid to the Fermi liquid in the whole doping region.
Therefore there is a point, at least, to distinguish
the two kinds of liquid, on both sides of which,
the physical properties should be fundamentally different. But where is it?
For now we cannot solve analytically the Hubbord model
in the strongly correlated system, we do not know the exact position of
this point and have to refer ourselves to experiment. 

In fact, Tallon and Loram {\it et al.} \normalsize in 1999 had been already
aware of the existence of such point and called it 
{\it quantum critical point} \normalsize (QCP)\cite{Tallon99}.
Subsequently they surveyed all the experimental results, which confirm
its existence: the physical properties on both sides of it are
fundamentally different\cite{Tallon01}.
Their works give us an explicit answer:
the critical point is in $p=0.19$ and it is $p=0.19$, not the
optimal doping point $p=0.16$, that divides the whole doping region
into two parts: the underdoped region with $ p<0.19$ and the overdoped
region with $p>0.19$.

The experiment by Uchida\cite{Uchida} also confirms that the carrier 
concentration in the underdoped region is proportional to the hole
concentration ($p$) and that in the overdoped region is proportional to
the electron concentration ($1-p$).

In a summary, the electrons in the underdoped region perform as
the non-Fermi liquid and the electrons in the overdoped region perform as 
the Fermi liquid. The former are localized by the on-site Coulomb repulsion
and pile up one by one, more similar to the solid; the latter are not 
localized even with the on-site Coulomb repulsion, more similar to the gas.
We will see it is the fundamental starting point
to discuss the exotic properties in the cuprate superconductors.
In the following we will discuss the doping effects on gap.

\section{Pseudogap and superconductivity gap}
\subsection{Pseudogap}

In the conventional superconductors, interaction by exchanging 
phonon induces electrons to pair as long as temperature 
drops low enough, which forms quasiparticles and decreases the 
energy of the system\cite{Cooper}. 
The expectation of the paired operators in the superconductivity state 
does not varnish and can be defined as the {\it energy gap\normalsize},
an energy scale to determine the critical temperature ($T_c$)\cite{BCS}.
In the common views, the quasiparticles arising from the paired electrons,
furthermore,
are in the boson mode and can form coherence at low enough temperature,
which implies there is a new energy scale independent of pairing.
It is considered that the pairing energy scale is lower than the 
coherence energy scale 
in the conventional superconductors and pairing once forms, 
quasiparticles instantly form coherence and superconductivity emerges;
but in the cuprate superconductors, the pairing energy scale, on the contrary,
is higher than the coherence energy scale, as a result, the gap in the normal state,
known as the {\it pseudogap\normalsize}, a precursor to the 
superconductivity gap, forms before coherence. All reviewed here
are called the {\it precursor pairing scenario\normalsize} in the 
literature\cite{Emery}.

It is not difficult, in fact, to check such scenario by experiment, 
or more explicitly, to see what happens when coherence is destroyed 
by strong external field. Strong magnetic fields experiment\cite{Zheng_exp}
tells us the pseudogap still exists below the critical temperature ($T_c$) in the 
underdoped region, when even the superconductivity gap, or the coherence,
is violated; there is no any gap, moreover, observed in the
overdoped region in the same condition, totally different from the 
precursor pairing scenario. 

What is the physical reason for the pseudogap?
We will see in the following that localization is responsible for the pseudogap.

The electons in the underdoped region, due to the on-site Coulomb repulsion,
are localized, or partly localized at
least, as discussed above. To decrease the energy of the system 
with superexchange interaction, two adjacent electrons overlap by
resonating to form the singlet, very similar with Anderson's
RVB scenario\cite{Anderson1987}. But it is only the adjacent completely-localized
electrons, the electrons that cannot hop, not all the electrons,
that form the singlet, which is a little different from
the RVB scenario or the precursor pairing scenario. 

Pairing makes the paired electrons inseparable and excludes the holes, 
which only hop around them. That is to say, the paired electrons are 
{\it frozen\normalsize}. Localization sustains the stability of pairing,
and in return, pairing strengthens localization. Were electrons itinerant
or movable, they would move apart and be free of the previous superexchange 
interaction, which is fatal to the adjacent pairing.

But not all the electrons are localized.
A hole not only provides the electrons around it opportunities to hop, 
but is also destructive to their localization, 
and, furthermore, to their pairing.
It manifests the carriers do not pair and 
the pseudogap paring is not the precursor to the superconductivity pairing.

In a summary, the pseudogap, or the normal-state gap, is only related to
the localized Fermi liquid
in the underdoped region and is not found in the Fermi liquid in the 
overdoped region, as expected. In the copper oxide planes all the paired electrons
in the pseudogap state form an {\it electron glacier\normalsize} with the help of localization, 
in which the unpaired electrons flow across the holes. Doping increases
the holes, which is destructive to localization, and more to 
the electron ice. When the doping concentration $ p=0.19$, 
the electron glacier avalanches 
and the electrons' flow is free of the constraint from the holes. The electrons
begin to perform as the Fermi liquid. 

In the Fermi liquid such mechanism cannot apply, for the electrons are
no longer localized but itinerant in the lattice and cannot pair respectively via 
the superexchange interaction, or even they pair, they can easily fall apart.
It can be understood by considering only two electrons in the lattice: 
they, due to the principle of uncertainty, exist in any possible
lattice and obtain very low possibility to pair in the neighborhood. 
To sustain such pairing, a collective mechanism is needed, which is 
{\it coherence\normalsize}.

\subsection{Coherence and superconductivity gap}
Coherence originates from the identity of the quasiparticles,
which consist of even fermions and are in boson mode.
The coherence-induced pairing, instead of
existing in partial electrons, applies to all electrons.
The electrons in coherence no longer move solely but participate
the collective resonance mode as quasiparticles.
The emergence of coherence implies electrons are in superconductivity.

Coherence in superconductivity shares the striking similarity
with that in the Bose-Einstein condensation (BEC),
but also distinguishes itself from the latter,
which supports all the peculiarities shown in the cuprate superconductors
and also misleads the physicist. 

We discuss the distinction between them first.

The chemical potential of the quasiparticles vanishes because
the number of the excited quasiparticles is not fixed. 
There is no constraint, therefore, from the density of particles,
which gives the relation between the critical temperature 
and the particle density in BEC. As a result, {\it all the energy 
scale is included in the superconductivity gap and there is no 
independent energy scale corresponding to coherence. }\normalsize
Though pairing of the carriers and coherence are considered as two
main characteristics of 
superconductivity, it is that one implies two and two merges into one:

 {\it The pairing of carriers implies coherence, and vice versa}.\normalsize

Like the role of localization for the pseudogap
in the underdoped region, it is coherence that makes paired 
carriers move together in the superconductivity state
and sustains the stability of pairing, whether in the underdoped region
or in the overdoped region.
The pairing of carriers provides the prerequisite to coherence
and in return, coherence keeps pairing stable;
they are interdependent and indispensable. 
Once coherence is violated, the pairing of the carriers, whether in the 
underdoped region or in the overdoped region,
disappears, so does superconductivity and the superconductivity
gap\cite{Zheng_exp}.

In the overdoped region, the electrons as carriers form pairing
and the superconductivity gap, the only candidate for the gap.
But in the underdoped
region, it becomes a little complicated. The paired carriers in 
the superconductivity state cover both the pairs with pseudogap
and the pairs which are single in the normal state, which coexist
and do not compete with other. Two kinds of quasiparticles participate
coherence together, which manifests the quasiparticles
are identical in the pairing pattern, as a result of which, the two gaps
ought to choose the same wave function to pair and share the same
physical properties,
as suggested by the experiments\cite{Tsuei,STM,Nernst}.
\subsection{Coherence-induced phonon}
Although the superconductivity in the overdoped region originates from
the Fermi liquid, the same background as that in the conventional
metals, it identifies itself with new features, as a result of
fitting itself the special condition in the cuprate superconductors.

The quasiparticles, or the paired electrons,
even in the Fermi liquid, align in the different lattices and 
transfer the variation of phase, or the thermal excitation in 
a new way. Suppose an excitation off the node arises from some
quasiparticle, it transfers to another quasiparticle by the 
identity of boson, and so on, which, sometimes called the
{\it phase fluctuation\normalsize} in the literature\cite{Emery},
gives an interpretation of “collective”.

The phase fluctuation, in nature, is a new kind of quasiparticle in 
superconductivity, very similar with the {\it phonon\normalsize}
in BEC, or in superfluidity\cite{Landau,Leggett}, 
The coherence-induced phonon, responsible 
for the phonon mode shown in the ARPRES experiments\cite{Shen},
dominates the thermal 
excitation around the node, where the energy of the paired
quasiparticle varnishes, 
though there is no explicit electron-phonon interaction
in the cuprate superconductors. The dispersion relation of the 
phonon in superconductivity is just that of the paired quasiparticle, for the energy
of the paired quasiparticles in the node
is exactly zero. The anisotropy of the gap implies the anisotropy
of phonon.

Consequently, we have to revise the traditional
BCS theory to apply in the cuprate superconductors, mainly
replacing the fermion excitation of quasiparticles with 
the excitation of phonon, a kind of boson.

Numerical result in rough approximation demonstrates the universal
relation between the critical temperature ($T_c$) and the 
superconductivity gap ($\Delta _0$),
$ \frac{2\Delta _0 }{ k_{\rm B} T_c } = 4.3$, 
does not exist, which is substituted by more complicated relationship.

\subsection{Amplitude of the gap}
The pairing, whether in the normal state or in the superconductivity 
state, originates from the resonating between the two adjacent electrons.
The amplitude of paring characterizes the degree of resonating, 
or more intuitively, the degree of overlapping.

In the pseudogap state, the less the holes, the more crowded the 
electrons, the more possibility to overlap and the larger the pseudogap;
in the superconductivity state, the less the carriers, the more 
fragile coherence, the less possibility to overlap and the smaller
superconductivity gap.

The crowdedness in the underdoped region does not change with temperature, therefore, nor
does the amplitude of the pseudogap; but in the overdoped region,
it becomes a little different: the thermal excitation leads to
the excitation of phonon, which does not affect pairing,
therefore, the amplitude of the superconductivity gap, likewise,
does not change with temperature. All in all, the amplitude of 
gap, no matter which kind of gap, does not change with temperature,
as suggested by the experiment\cite{STM}.

Compared with the pseudogap in the normal state, the superconductivity
gap, a product of the collective condensation, is more fragile and
therefore, has smaller gap amplitude and lower critical temperature.
\section{The normal state}
\subsection{Normal state in the overdoped region}
In the overdoped region the electrons are in the state of 
the antiferromagnetic Fermi liquid(AFL), in which the effect
of the spin density fluctuations dominates the properties 
in the thermal equilibrium state and is especially responsible 
for the linear resistivity\cite{Millis}.

In transportation, electrons avoid the spin-flip process
to decrease the resistivity by exchanging the momentum only, leaving
the spin unchanged in the scattering, or keeping the original
spin order, which, we name as {\it superexchange transportation}.
\normalsize 

\subsection{Normal state in the underdoped region}
Though the spin density fluctuations still take effects in the
underdoped region, the localization of the electrons plays
a decisive role and generates new features, one of which, the
pseudogap, for example, has been discussed above, dividing the 
normal state into two parts:
\subsubsection{Normal state without the pseudogap}
The electrons in the thermal equilibrium without external field
can be in spin-disorder, but in transportation
“the hopping of an isolated hole
leaves a line of misaligned spins in its wake”\cite{Orenstein},
which, known as {\it frustration}\normalsize, produces a great 
many of spin-flip processes and consumes additional unnecessary
energy.
Frustration, therefore, needs to be avoided as possible, thus 
the charge order and spin order is the only alternative. 
In some special doping point, for example, $p=\frac{1}{8}$, 
frustration varnishes and electrons flow orderly between the
domain walls, which is the so-called {\it stripe}\normalsize,
as discovered in Ref\cite{Tranquada}.
In the very low doping area, holes are so rare that it hard to 
form spin order, as a result, frustration dominates transportation.

To obtained the charge and spin order, massive spin-flip processes
are inevitable at the very beginning of transportation, 
which leads to that the initial resistivity
is by far larger than the static resistivity.

Had the charge and spin order been determined, the spin density
fluctuations would dominate the properties of the electrons. 
This is why the resistivity in the normal state without gap 
shares the similar dependence on temperature, whether 
in the underdoped region or in the overdoped region. 

\subsubsection{Normal state with the pseudogap}
The paired electrons in the normal state, unlike those in 
the superconductivity state, do not form a 
collective behaviour. The excitation from the paired quasiparticles
melts them into liquid, or depair them. As temperature increases, 
the quasiparticles varnish gradually. Though the evolution of the 
pseudogap is different
from the superconductivity gap, the critical temperature ($T^\ast$)
can be calculated by similar method, for the melted electrons
are equivalent to the mixture of the quasiparticles in the ground
state and  in the excited state.

In a summary, the pseudogap is like a floating ice on the Fermi
sea in the momentum space and melts gradually till the critical temperature ($T^\ast$).

When the electrons form pseudogap, they are frozen and do not
participate transportation, they also decouple from 
other electrons. The spin density fluctuation, therefore, 
only emerges from the unpaired electrons. 
As temperature increases, the number of unpaired electrons 
increases, which makes the resistivity deviate from the linear 
behaviour\cite{Timusk}.
\section{Summary}
After above discussion, we summarize our arguments in the following:
\begin{enumerate}
	\item The hole concentration $p =0.19$ is a quantum
		critical point, dividing the whole doping region into
		two parts: the underdoped region ($p < 0.19 $) and 
		the overdoped region ($ p> 0.19 $). The electrons 
		 in the former are localized and form the localized Fermi
		liquid, a kind of the non-Fermi liquid, where the carriers
		are holes; the electrons in the latter are itinerant
		and form the Fermi liquid, where the carriers are
		the electrons themselves.
	\item Gap is a way to decrease the energy of the system. But
		 different from the pairing in the 
		conventional superconductors, the pairing in the
		cuprate superconductors only arises from two adjacent electrons
		and decouple them from the others.

		The normal state gap, or the pseudogap, a gap without
		coherence, only originates from the localized Fermi liquid for 
		localization is a prerequisite to keeping no-coherence 
		pairing stable. The paired quasiparticles in the normal state,
		instead of varnishing suddenly in the critical point,
		 decrease gradually and varnish finally
		in the critical temperature.
	\item 
		Superconductivity is the condensation of the
		adjacently-paired carriers, whether in the underdoped region or 
		in the overdoped region. The pairing of the carriers immediately
		implies coherence. If coherence is violated, the pairing of
		carriers varnishes.

		Coherence induces a new quasiparticle {\it phonon\normalsize},
		a kind of boson, which is responsible for the 
		excitation of the quasiparticles around the node.
		The boson distribution, instead of the fermion distribution,
		therefore, is applied in the calculation.
	\item The pseudogap coexists with the superconductivity gap in 
		the superconductivity state of the underdoped region.
		Once coherence is violated, only the pseudogap is left.
	\item A new kind of spin order,  stripes,
		forms in transportation in the underdoped 
		region to avoid frustration. 
\end{enumerate}
There are still a lot of issues without definite answers:
\begin{enumerate}
	\item The properties in the normal state need quantitative results
		to fit the experimental results, especially in the 
		underdoped region.
	\item We believe the pseudogap in other compound, for example, 
		the manganeses\cite{Shen_pg}, 
		should share the same physical essence as in the
		cuprate superconductors, but details need further examination.
	\item When stripe forms, the $2$-dimension superconductivity
		turns into $1$-dimension superconductivity. 
		Its details also need further examination.
\end{enumerate}

We once expect our suggestion can cover all the issues
in the cuprate superconductors and construct 
an overwhelming theory. But its complexity is beyond our own ability. 
Even so, we believe it will provide a promising understanding on HTSC.
No river, after all, can support a large ship in its origin.

\section*{Acknowledgement}
The author is grateful to Dr. C.-G. Ma and F.-W. Shu for help discussions.
This work was supported in part by Natural Science Foundation of ChongQing 
under the Grant 2009JJ1304.

\end{document}